\documentstyle[12pt,epsfig]{article}                                                       
                                                       
\newlength{\dinwidth}                                                       
\newlength{\dinmargin}                                                       
\def\lapproxeq{\lower .7ex\hbox{$\;\stackrel{\textstyle                                                       
<}{\sim}\;$}}                                                       
\def\gapproxeq{\lower .7ex\hbox{$\;\stackrel{\textstyle                                                       
>}{\sim}\;$}}                                                       
\def\be{\begin{equation}}                                                       
\def\ee{\end{equation}}                                                       
\def\bea{\begin{eqnarray}}                                                       
\def\eea{\end{eqnarray}}                                                       
                                                       
\def\funp{{I\!\!P}}                       
                                                       
\begin{document}                                                       
\titlepage                                                       
\begin{flushright}                                                       
IPPP/00/09 \\ 
DTP/00/72 \\                                                       
23 November 2000 \\                                                       
\end{flushright}                                                       
                                                       
\vspace*{2cm}                                                       
  
\renewcommand{\thefootnote}{\fnsymbol{footnote}}                                                     
\begin{center}                                                       
{\Large \bf Soft diffraction at the LHC and properties} \\

\vspace*{0.5cm}
{\Large \bf of the Pomeron\footnote[2]{Based on a talk by M.G. Ryskin at Diffraction 2000, 
Cetraro, Italy, September 2000}} \\                                                       

\vspace*{1cm}                                                       
V.A. Khoze$^a$, A.D. Martin$^a$ and M.G. Ryskin$^{a,b}$ \\                                                       
                                                      
\vspace*{0.5cm}                                                       
$^a$ Department of Physics and Institute for Particle Physics Phenomenology, University of  
Durham, Durham, DH1 3LE \\    
$^b$ Petersburg Nuclear Physics Institute, Gatchina, St.~Petersburg, 188300, Russia               
\end{center}                                                       
                                                       
\vspace*{2cm}                                                       
                                                       
\begin{abstract}                                                       
We briefly describe a model for soft diffraction.  The model description of ISR, $Sp\bar{p}S$  
and Tevatron data, and the predictions of the total, elastic, single and double diffractive cross  
sections at the LHC can be found in the original paper.  Here we address issues raised at  
Diffraction 2000, in particular concerning the nature of the resulting Pomeron trajectory. 
\end{abstract}                                             
        
\newpage  
\renewcommand{\thefootnote}{\arabic{footnote}}                                                       
            
\section{Model for soft diffraction}      

Here we discuss a recent description \cite{SOFT} of soft diffraction in high energy $pp$ (or  
$p\bar{p}$) collisions which embodies 
\begin{itemize} 
\item[(i)] {\it pion-loop} insertions in the bare Pomeron pole, which represent the nearest  
singularity generated by $t$-channel unitarity, 
\item[(ii)] a {\it two-channel eikonal} which incorporates the Pomeron cuts generated by  
elastic and quasi-elastic (with $N^*$ intermediate states) $s$-channel unitarity, 
\item[(iii)] high-mass {\it diffractive dissociation}. 
\end{itemize} 
 
To implement (i), we follow Anselm and Gribov \cite{AG} and write the Pomeron trajectory 
\begin{equation} 
\label{eq:a1} 
\alpha_\funp (t) \; = \; \alpha (0) \: + \: \alpha^\prime t \: + \: \frac{\beta_\pi^2 m_\pi^2}{32  
\pi^3} \: h (t), 
\end{equation} 
where $\alpha (0) + \alpha^\prime t$ is the {\it bare} trajectory and where the pion-loop  
insertions are given by 
\begin{equation} 
\label{eq:a2} 
h (t) \; = \; - \frac{4}{\tau} \: F_\pi^2 (t) \left [ 2 \tau \: - \: (1 + \tau)^{3/2} \: \ln \left (  
\frac{\sqrt{1 + \tau} + 1}{\sqrt{1 + \tau} - 1} \right ) \: + \: \ln \frac{\mu^2}{m_\pi^2} \right  
], 
\end{equation} 
with $\tau = 4 m_\pi^2/| t |$ and $\mu = 1~{\rm GeV}$.  The coefficient $\beta_\pi^2$  
specifies the $\pi\pi$ total cross section and $F_\pi (t)$ is the form factor of the pion-Pomeron  
vertex.  The coefficient of $h (t)$ in (\ref{eq:a1}) is small but, due to the tiny scale $m_\pi$,  
the $t$ dependence of $h (t)$ is steep and non-linear.  In this way we correctly reproduce the  
behaviour of the diffractive amplitude in the {\it peripheral} region. 
 
Note that $h (t)$ of (\ref{eq:a2}) has been renormalized \cite{AG} such that 
\begin{equation} 
\label{eq:a3} 
h (t) \; = \; h_\pi (t) \: - \: h_\pi (t = 0), 
\end{equation} 
where $h_\pi (t)$ denotes the full pion-loop contribution.  The value of $h_\pi (0)$ is  
determined by large virtualities of the pion, which are controlled by the scale $\mu$.  For this  
reason the contribution, 
\begin{equation} 
\label{eq:a4} 
\beta_\pi^2 m_\pi^2 \: h_\pi (0)/32 \pi^3 \: \sim \: 0.1, 
\end{equation} 
has been included in $\alpha (0)$, which describes the {\it small size} $(\sim 1/\mu)$  
component of the Pomeron. 
 
\section{The results} 
 
The above model describes the data on the total and differential elastic cross section  
throughout the ISR-Tevatron energy interval, see \cite{SOFT}.  Surprisingly, we found the  
bare Pomeron parameters to be 
\begin{equation} 
\label{eq:a5} 
\Delta \; \equiv \; \alpha (0) \: - \: 1 \: \simeq \: 0.10, \quad\quad \alpha^\prime \; = \; 0. 
\end{equation} 
On the other hand it is known that the same data can be described by a simple effective  
Pomeron pole with \cite{DL} 
\begin{equation} 
\label{eq:a6} 
\alpha_\funp^{\rm eff} (t) \; = \; 1.08 \: + \: 0.25~t. 
\end{equation} 
In our approach the shrinkage of the diffraction cone comes not from the bare pole  
($\alpha^\prime = 0$), but has components from the three ingredients, (i)--(iii), of the model.   
That is, in the ISR-Tevatron energy range
\begin{equation} 
\label{eq:a7} 
{\rm \lq\lq} \alpha^\prime_{\rm eff} \mbox{\rq\rq} \; = \; (0.034 \: + \: 0.15 \: + \: 0.066)~{\rm 
GeV}^{-2} 
\end{equation} 
from the $\pi$-loop, $s$-channel eikonalization and diffractive dissociation respectively.   
Moreover eikonal rescattering suppresses the growth of the cross section and so $\Delta  
\simeq 0.10 > \Delta_{\rm eff} \simeq 0.08$.  Note that at lower energies the fixed target data require 
$\alpha_{\rm eff}^\prime = 0.14~{\rm GeV}^{-2}$ \cite{NA48}, which is consistent since as the energy 
decreases the effect of the eikonal and higher mass diffractive dissociation is less than in 
(\ref{eq:a7}).
 
It is important to discuss the results for the behaviour of the local slope of the forward elastic  
peak 
\begin{equation} 
\label{eq:a8} 
B (t) \; = \; d \: \ln (d \sigma_{\rm el}/dt)  / dt. 
\end{equation} 
The $\pi$-loop insertions lead to a contribution to $B (t)$ which increases as $|t| \rightarrow  
0$ (and behaves as $\ln s$).  On the other hand the eikonal absorptive corrections lead to a  
dip in $d \sigma_{\rm el}/dt$ whose position moves to smaller $|t|$ as $\sqrt{s}$ increases,  
which causes $B (t)$ to decrease as $|t| \rightarrow 0$.  Fortunately just at the LHC energy,  
these two effects compensate each other and $B$ is essentially independent of $t$ for $|t| <  
0.1~{\rm GeV}^2$ \cite{SOFT}. 
 
The solid and dashed curves in Fig.~1\footnote{We thank A. Kaidalov for emphasizing the  
value of this type of plot.} show, respectively, the energy behaviour of the slope $B (0)$ and  
$\sigma_{\rm tot}$ (for the \lq maximal\rq\ model of \cite{SOFT}).  We plot $\sigma_{\rm  
tot}$ in fm$^2$ and $B$ in GeV$^{-2}$, since in these units the asymptotic black-disc limit  
takes, to a good approximation, the simple form $B/\sigma_{\rm tot} \rightarrow 1$.  For  
comparison the dotted lines show the simple effective Pomeron results of \cite{DL}.  It is  
clearly seen that, while the simple and detailed models give similar values of $\sigma_{\rm  
tot}$ up to LHC energies, the predictions for the slope $B (0)$ already differ significantly. 
 
\begin{figure} 
\begin{center} 
\psfig{figure=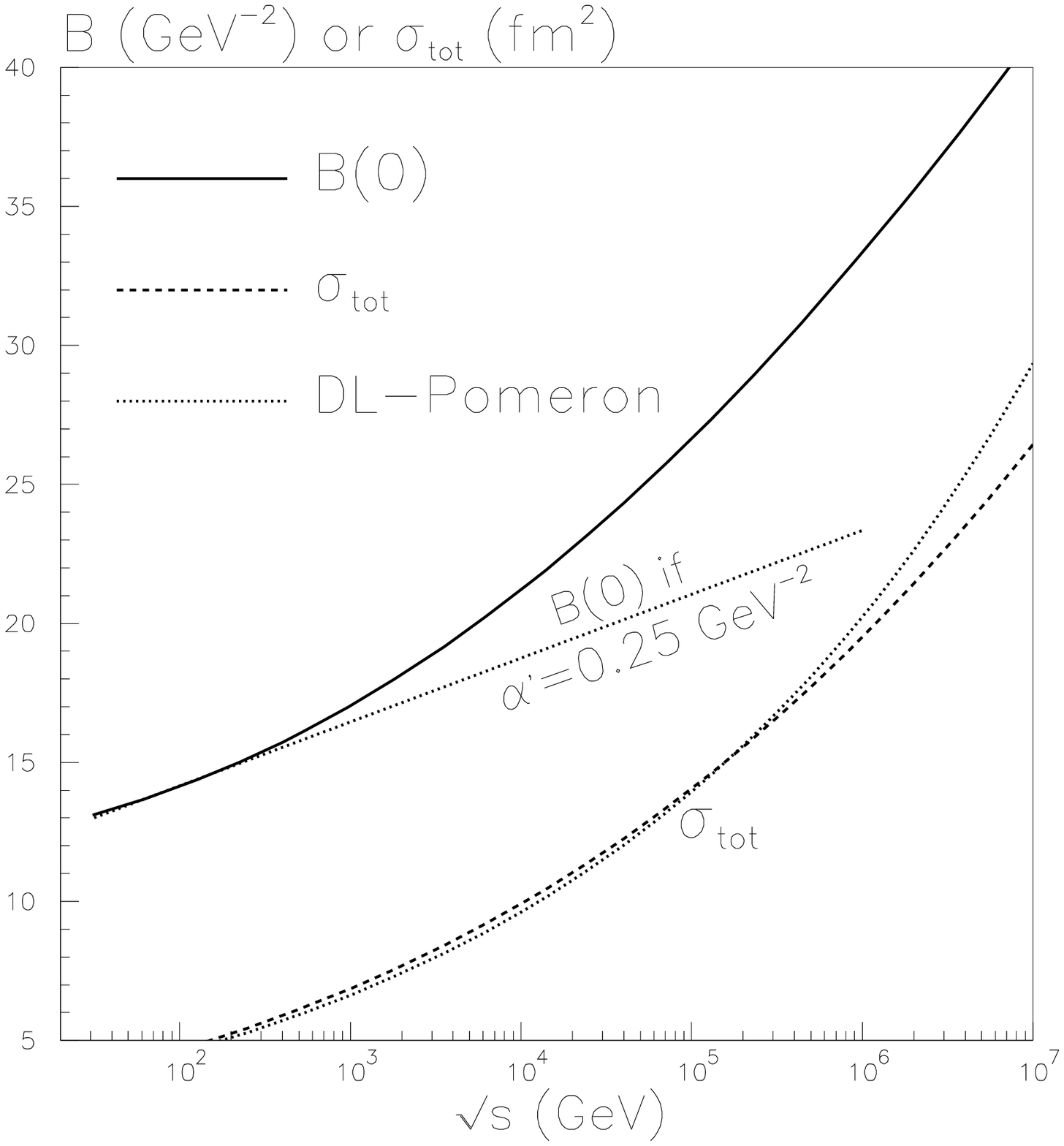,width=8cm} 
\end{center} 
\caption{The energy behaviour of the forward elastic slope and the total cross section,  
compared with that obtained from the simple effective Pomeron pole of (\ref{eq:a6})  
\cite{DL}. 
\label{fig:Fig1} 
} 
\end{figure} 
 
Since our model \cite{SOFT} embodies all the main features of soft diffraction we expect it  
to give reliable predictions for the {\it survival probability} $S^2$ of the rapidity gaps against  
population by secondary hadrons from the underlying event, that is hadrons originating from  
soft rescattering.  In particular we predict $S^2 = 0.10 (0.06)$ for single diffractive events  
and $S^2 = 0.05 (0.02)$ for exclusive Higgs boson production $pp \rightarrow p + H + p$ at  
Tevatron (LHC) energies. 
 
\section{Absorptive corrections are greater than $2 \sigma_{\rm el}/\sigma_{\rm tot}$} 
 
It is informative to recall the origin of the eikonal rescattering corrections, ingredient (ii) of  
our model.  Multi-Pomeron vertices are complicated objects which are not closely related to  
$\sigma_{\rm el}$, as represented by the two-Pomeron exchange diagram of Fig.~2(a).   
Indeed Fig.~2(a) vanishes as $s \rightarrow \infty$ due to the cancellation between the cut  
shown by line 1 with those shown by lines 2 and 3 \cite{AFS}.  Rather the true two-Pomeron  
vertex is given by the \lq\lq Mandelstam crossed\rq\rq\ diagram of Fig.~2(b) \cite{MS} and,  
moreover, it cannot be too small.  The reason is that the inelastic cuts 2, 3 of Fig.~2(a) give a  
{\it negative} cross section $\sigma_{SD}$ for single diffractive dissociation (of the upper  
hadron) which is exactly cancelled by the positive elastic contribution of cut 1.  So to obtain a  
positive $\sigma_{\rm SD}$ we require the contribution Fig.~2(b) to be larger than  
$\sigma_{\rm el}$ (cut 1 of Fig.~2(a)).  In general the multi-Pomeron vertex corresponds to the 
complete set of \lq 4-leg\rq\ proton-Pomeron diagrams which possess both $s$- and $u$-channel cuts 
simultaneously.  Fig.~2(b) is the simplest such diagram.  Moreover it gives the leading order 
$(\log s)$ contribution, and illustrates the main properties of the two Pomeron vertex.  
An analogous result follows for multi-Pomeron  
exchanges.  We conclude that the absorptive correction due to $s$-channel rescattering,  
Fig.~2(b), cannot be smaller than that given by the elastic eikonal \cite{AGK}\footnote{See \cite{BR} for  
a more detailed discussion of the AFS cancellation in the case of QCD.}; that is $\sigma_{\rm  
el}/\sigma_{\rm tot}$ is the minimal correction to the elastic amplitude and $2 \sigma_{\rm  
el}/\sigma_{\rm tot}$ is the correction for the inelastic cross section. 
 
\begin{figure} 
\begin{center} 
\psfig{figure=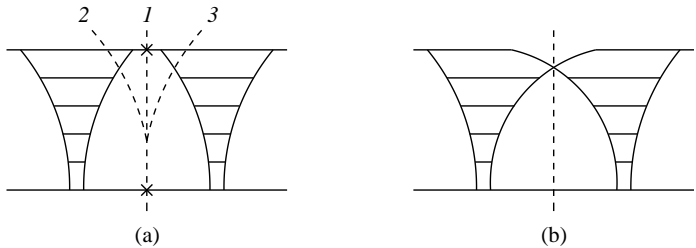,width=10cm} 
\end{center} 
\caption{The Feynman diagram for the two-Pomeron cut, drawn in such a way to reflect  
the space-time picture of the interaction, for (a) AFS elastic rescattering \cite{AFS} and (b)  
Mandelstam crossed diagram \cite{MS}. 
\label{fig:Fig2} 
} 
\end{figure}

\section{The heavy Pomeron and glueballs} 
 
To describe the data, our {\it bare} Pomeron, in the {\it absence} of pion-loop  
insertions\footnote{The remaining loop insertions, including the Pomeron-loop  
renormalisation, are embodied in the parameters of the bare pole.}, has 
\begin{equation} 
\label{eq:a9} 
\alpha (0) \; \simeq \; 1, \quad\quad \alpha^\prime \; \simeq \; 0, 
\end{equation} 
see (\ref{eq:a5}) with (\ref{eq:a4}) subtracted.  In terms of QCD, does this mean the bare  
Pomeron is described by two-gluon exchange $\grave{a}$ {\it la} Low and Nussinov  
\cite{LN}?  In fact, the situation is more complicated, since at scale $\mu \sim 1~{\rm GeV}$  
we have multi-gluon exchange and probably gluon saturation, as in the black-disc limit. 
 
Rather, $\alpha^\prime = 0$, means that the bare Pomeron has no $b_t$ dependence and lives  
in $1 + 1$ dimensions, where the Froissant bound corresponds to a fixed bare pole with  
$\alpha = 1$.  This picture is very similar to the \lq\lq heavy\rq\rq\ Pomeron proposed by  
Gribov \cite{GR}.  At first sight, for a fixed pole $\alpha = 1$, it is impossible to obtain a  
mass spectrum for glueballs which are believed to lie on the Pomeron trajectory.  Indeed in  
pure gluodynamics (without quarks and pions) we would then have no glueballs.  This is in  
accord with Gribov's claim that there is no confinement in a World without light quarks  
\cite{G2}.   
 
However, we have to account for the $t$-channel unitarity iterations when quarks (and pions,  
which are the chiral phase of the light quark fields) exist.  As  
seen above, the pion-loop insertion shifts $\alpha (0)$ and produces a non-zero  
$\alpha_\funp^\prime$.  Next, as emphasized in \cite{KS}, there is an important mixing  
between the Pomeron and the bare $f (u\bar{u} + d\bar{d})~[f^\prime (s\bar{s})]$ meson  
trajectories in the time-like $t \sim 0.5-1~{\rm GeV}^2$ domain.  After mixing, the leading  
Pomeron trajectory follows the original $f$-trajectory in the time-like domain\footnote{This  
is not in contradiction with our much flatter Pomeron trajectory in the space-like domain,  
where $f$-Pomeron mixing is negligible and $\alpha_\funp^\prime$ is small.}.  Therefore the  
expected glueball masses should be similar to those predicted in \cite{KS} (which, in turn,  
agrees with lattice expectations). 
 
\section*{Acknowledgements} 
 
We thank K. Goulianos, A.B. Kaidalov and P.V. Landshoff for useful discussions.

\end{document}